\documentstyle[12pt,epsf,epsfig]{article}

\newcommand{\beq}{ \begin{eqnarray} }
\newcommand{\eeq}{ \end{eqnarray} }
\newcommand{\beqstar}{ \begin{eqnarray*} }
\newcommand{\eeqstar}{ \end{eqnarray*} }

\newcommand{\lsim}{ \mathop{}_{\textstyle \sim}^{\textstyle <} }

\begin{document}
\baselineskip 0.7cm

\begin{titlepage}

\begin{center}

\hfill ICRR-Report-496-2002-14\\
\hfill DPNU-02-40\\
\hfill \today

{\large 
GUT relation in Neutrino-Induced Flavor Physics\\
in  SUSY SU(5) GUT 
}
\vspace{1cm}

{\bf Junji Hisano}$^{1}$  and 
{\bf Yasuhiro Shimizu}$^{2}$
\vskip 0.15in
{\it
$^1${ICRR, University of Tokyo, Kashiwa 277-8582, Japan }\\
$^2${Department of Physics, Nagoya University, Nagoya 464-8692, Japan}\\
}
\vskip 0.5in

\abstract{
In the SUSY SU(5) GUT with the right-handed neutrinos, the neutrino
Yukawa coupling induces the flavor-violating mass terms for the
right-handed down squarks and left-handed sleptons, and they lead
to imprint on the flavor physics in the hadronic and leptonic sectors.
In this paper we study CP violation in $B_d^0$ and $B_s^0$, which is
induced by the tau-neutrino Yukawa coupling. In particular, we notice the
correlation of the CP asymmetries in $B_d^0\rightarrow \phi K_s$
($S_{\phi K_s}$) and $B_s^0\rightarrow J/\psi
\phi$ ($S_{J/\psi \phi}$) with other observables, such as
$Br(\tau\rightarrow\mu\gamma)$, the muon EDM ($d_\mu$), and the
mixing-induced CP asymmetry in $B_d^0\rightarrow M_s^0\gamma$ ($A_{bsg}$),
since they have a common origin of the CP and flavor violation. The
measurements are expected to play a significant role for consistency
check of the SUSY SU(5) GUT. The hierarchical right-handed neutrino
masses make the correlation better. We find that the current 
bound on $Br(\tau\rightarrow\mu\gamma)$ constrains the allowed region 
for $S_{\phi K_s}$ in this model.
}
\end{center}
\end{titlepage}
\setcounter{footnote}{0}

The Cabibbo-Kobayashi-Maskawa (CKM) paradigm works well to explain the
observed flavor- and CP- violating processes in the hadronic sector. It
has passed the non-trivial tests using $K$ and $B$ mesons. The recent
observations for $\sin2 \beta$ in the Belle \cite{belle} and Babar
\cite{babar} are consistent with other processes, such as the
$K^0$-$\bar{K}^0$ mixing, and the unitarity triangle looks close in
the $\rho$-$\eta$ plane. The CKM dominance in the flavor- and CP-
violating processes in the hadronic sector is almost established now.
However, still there are rooms open to the new physics, and the clue
may be found in the $B$ factories or the hadron colliders in near
future.

The neutrino-oscillation experiments
\cite{skatm}\cite{Ahmad:2002jz}\cite{kamland} show that the lepton
sector has large flavor violation, contrary to the quark sector in the
standard model (SM). In the supersymmetric SU(5) grand unified model
(SUSY SU(5) GUT), the neutrino sector supplies a new source of the
flavor and CP violation in the hadron physics
\cite{moroi}\cite{susygutb}\cite{gotoshimizu}\cite{gotoshimizu2}\cite{Moroi:2000tk}\cite{hlm},
when the SUSY breaking terms in the minimal SUSY SM (MSSM) are
generated above the GUT scale.  In the supersymmetric seesaw model
\cite{seesaw} the flavor violation in the neutrino sector  may 
lead to  charged lepton-flavor violating processes, such as
$\mu \rightarrow e \gamma$ or $\tau \rightarrow \mu \gamma$
\cite{bm}\cite{lfv}\cite{hn}\cite{ci}, since the radiative correction by the
neutrino Yukawa coupling leads to the non-vanishing
lepton-flavor violating mass terms for left-handed sleptons. In the
SUSY SU(5) GUT the left-handed leptons are embedded in common
multiplets with the right-handed down quarks, and this means that the
right-handed down squarks may also have the flavor-violating mass
terms.

This may lead to the sizable effects to the $K$ and $B$ meson
systems. In particular, if the tau-neutrino Yukawa coupling is of the
order of one, similar to the top-quark one, the $b$-$s$ transition in
the right-handed current is important since the the
atmospheric-neutrino result suggests the large mixing angle between
the second and third generations.  Recently, it is announced that the
CP asymmetry in $B_d^0\rightarrow \phi K_s$ is 2.7$\sigma$ deviated from
one in $B_d^0
\rightarrow J/\psi K_s$ in Belle and Barbar \cite{BKphi}. The
$b\rightarrow s \bar{s} s$ transition is sensitive to the new physics
since it is induced at one-loop level
\cite{gw}. Thus, the deviation might be a
signature of the effect of the neutrino Yukawa coupling 
\cite{hlm}\cite{phikrecent}.

In this paper we study CP violation in transition between the second
and third generations in the hadronic sector, which is induced by the
tau-neutrino Yukawa coupling in the SUSY SU(5) GUT. In particular, we
notice the correlation of the CP asymmetries in $B_d^0\rightarrow \phi
K_s$ ($S_{\phi K_s}$) and $B_s^0\rightarrow J/\psi \phi$ ($S_{J/\psi
\phi}$) with other observables, such as
$Br(\tau\rightarrow\mu\gamma)$, the muon EDM ($d_\mu$), and the
mixing-induced CP asymmetry in $B^0_d\rightarrow M^0_s\gamma$ ($A_{bsg}$),
since they have a common origin of the CP and flavor violation. The
measurements are expected to play a significant role for consistency
check of the SUSY SU(5) GUT.

First, we review the SUSY SU(5) GUT with right-handed neutrinos from a
viewpoint of the flavor physics. The Yukawa coupling for quarks and
leptons and the Majorana mass for the right-handed neutrinos in this
model is given as
\begin{eqnarray}
W&=& 
\frac14 f_{ij}^{u} \Psi_i \Psi_j H 
+\sqrt{2} f_{ij}^{d} \Psi_i \Phi_j \overline{H}
+f_{ij}^{\nu} \Phi_i \overline{N}_j {H}
+M_{ij} \overline{N}_i \overline{N}_j,
\label{superp_gut}
\end{eqnarray}
where $\Psi$ and $\Phi$ are {\bf 10}- and {$\bf \bar{5}$}-dimensional
multiplets, respectively, and $\overline{N}$ is the right-handed
neutrinos.  $H$ ($\overline{H}$) is {\bf 5}- ({$\bf \bar{5}$}-)
dimensional Higgs multiplets.  

After removing the unphysical degrees of freedom, the Yukawa coupling
constants and Majorana masses in Eq.~(\ref{superp_gut}) are given as
follows,
\begin{eqnarray}
f^u_{ij} &=& 
V_{ki} f_{u_k} {\rm e}^{i \varphi_{u_k}}V_{kj}, \\
f^d_{ij} &=& f_{d_i} \delta_{ij},\\  
f^\nu_{ij} &=& {\rm e}^{i \varphi_{d_i}} 
U^\star_{ij} f_{\nu_j} ,
\\
M_{ij} &=&  {\rm e}^{i \varphi_{\nu_i}} W_{ik} 
             M_{N_k} {\rm e}^{2 i \overline{\varphi}_{\nu_k}} 
            W_{jk} {\rm e}^{i \varphi_{\nu_j}},  
\end{eqnarray}
where $\sum_i \varphi_{f_i} =0$ $(f=u,d,\nu)$. Each unitary matrices
$U$, $V$, and $W$ have only a phase. Here, $\varphi_{u}$ and
$\varphi_{d}$ are CP violating phases inherent in the SUSY SU(5)
GUT. The unitary matrix $V$ is the CKM matrix in the extension of the
SM to the SUSY SU(5) GUT. When the Majorana mass matrix for the
right-handed neutrinos is diagonal, $U$ is the the MNS matrix, since
the light-neutrino mass matrix is
\begin{eqnarray}
(m_\nu)_{ij}
&=&
(f^{\nu}(M^{-1}) f^{\nu T})_{ij} \langle H_f \rangle^2
\end{eqnarray}
where $H_f$ is a doublet Higgs in $H$.

The colored-Higgs multiplets $H_c$ and $\overline{H}_c$ are introduced
in $H$ and $\overline{H}$ as SU(5) partners of the Higgs doublets in
the MSSM, respectively\footnote{
While the proton decay by the dimension-five operator, which is
induced by the colored-Higgs exchange, is a serious problem in the
minimal SUSY SU(5) GUT \cite{dim5}, it depends on the structure of the
Higgs sector \cite{dim5sup}. In this paper, we ignore the proton decay
while we adopt the minimal structure of the Higgs sector.
}, and they have new flavor-violating interactions.
Eq.~(\ref{superp_gut}) is represent by the fields in the MSSM as
follows,
\begin{eqnarray}
W&=& W_{MSSM+\overline{N}} 
\nonumber\\
&&
+ \frac12 
V_{ki} f_{u_k} {\rm e}^{i \varphi_{u_k}}V_{kj} 
 Q_i Q_j H_c 
+  
f_{u_i} V_{ij} {\rm e}^{i \varphi_{d_j}}
\overline{U}_i \overline{E}_j H_c 
\nonumber\\
&&
+ 
f_{d_i} {\rm e}^{-i \varphi_{d_i}}
Q_i L_i \overline{H}_c 
+ 
{\rm e}^{-i \varphi_{u_i}}
V_{ij}^\star
f_{d_j} \overline{U}_i \overline{D}_j \overline{H}_c 
+ 
{\rm e}^{i \varphi_{d_i}}
U_{ij}^\star f_{\nu_j}
\overline{D}_i \overline{N}_j H_c.
\end{eqnarray}
Here, the superpotential in the MSSM with the right-handed neutrinos
is
\begin{eqnarray}
W_{MSSM+\overline{N}} &=&
V_{ji} f_{u_j}  Q_i \overline{U}_j H_f 
+
f_{d_i} Q_i \overline{D}_i \overline{H}_f
+
f_{d_i} L_i \overline{E}_i \overline{H}_f
\nonumber\\
&&+U^\star_{ij} f_{\nu_j} L_i \overline{N}_j H_f
+M_{ij} \overline{N}_i \overline{N}_j.
\label{superp_mssm}
\end{eqnarray}
The flavor-violating interactions absent in the MSSM emerge in the
SUSY SU(5) GUT due to existence of the colored-Higgs multiplets.

If the SUSY-breaking terms in the MSSM are generated by interactions
above the colored-Higgs mass, such as in the supergravity, the
sfermion mass terms may get sizable corrections by the colored-Higgs
interactions. In this paper we assume the minimal supergravity
scenario, that is, the scalar masses and the trilinear SUSY-breaking
terms at the reduced Planck mass scale ($M_G$) are given universally
by $m_0$ and $A_0$, respectively. In this case, the flavor-violating
SUSY-breaking mass terms at low energy are induced by the radiative
correction, and they are approximately given as
\begin{eqnarray}
(m_{{Q}}^2)_{ij}  &\simeq&-\frac{2}{(4\pi)^2} 
V_{ki}^\star
f_{u_k}^2
V_{kj} (3m_0^2+ A_0^2) (3 \log\frac{M_G}{M_{GUT}}
                        + \log\frac{M_{GUT}}{M_{SUSY}}),\nonumber\\
(m_{\overline{U}}^2)_{ij}  &\simeq&-\frac{4}{(4\pi)^2} 
{\rm e}^{i\varphi_{u_i}}V_{ik}
f_{d_k}^2
V_{jk}^\star {\rm e}^{-i\varphi_{u_j}} (3m_0^2+ A_0^2)
\log\frac{M_G}{M_{GUT}}, \nonumber\\
(m_{\overline{D}}^2)_{ij}  &\simeq&-\frac{2}{(4\pi)^2} 
{\rm e}^{-i\varphi_{d_i}}U_{ik}
f_{\nu_k}^2
U^\star_{jk} {\rm e}^{i\varphi_{d_j}} (3m_0^2+A_0^2) 
\log\frac{M_G}{M_{GUT}},\nonumber\\
(m_{L}^2)_{ij}  &\simeq&-\frac{2}{(4\pi)^2} 
U_{ik}
f_{\nu_k} {\rm e}^{i\varphi_{\nu_k}}
W_{kl}
W^\star_{ml}
{\rm e}^{-i\varphi_{\nu_m}} f_{\nu_m} 
U_{jm}^\star(3m_0^2+ A_0^2) 
\log\frac{M_G}{M_{N_l}},\nonumber\\
(m_{\overline{E}}^2)_{ij}  &\simeq&-\frac{6}{(4\pi)^2} 
{\rm e}^{-i\varphi_{d_i}} V^\star_{ki}
f_{u_k}^2
V_{kj} {\rm e}^{i\varphi_{d_j}} (3m_0^2+ A_0^2)
\log\frac{M_G}{M_{GUT}},
\label{sfermionmass}
\end{eqnarray}
where $i\ne j$. Here, $M_{GUT}$ and $M_{SUSY}$ are the GUT scale and
the SUSY-breaking scale in the MSSM, respectively. In the MSSM with
the right-handed neutrinos, the flavor-violating structures appear
only in the left-handed squark and left-handed slepton mass
matrices. On the other hand, in the SUSY SU(5) GUT, other sfermions
may also have sizable flavor violation. In particular, the CP violating
phases inherent in the SUSY SU(5) GUT appear in
$(m_{\overline{U}}^2)$, $(m_{\overline{D}}^2)$, and
$(m_{\overline{E}}^2)$ \cite{moroi}.

In this paper we assume the minimal structure for the Yukawa coupling
constants given in Eq.~(\ref{superp_gut}). The $b/\tau$ mass ratio may
be explained by it while the down-type quark and charged lepton masses
in the first and second generations are not compatible. The
modification of the Yukawa sector, such as introduction of the
higher-dimensional operators, the vector-like matters, or the
complicate Higgs structure, may change the low-energy prediction for
the flavor violation, especially, between the first and second
generations.  On the other hand, we concentrate on the transition
between the second and third generations in this paper. We assume that
the Yukawa coupling constants, including the neutrino one, are
hierarchical, and that the extra interaction gives negligible
contribution to the transition between the second and third generations.

The neutrino-induced off-diagonal terms for the sfermion masses are
$(m_{\overline{D}}^2)_{ij}$ and $(m_{L}^2)_{ij}$ ($i\ne j $).  Let us
demonstrate the good correlation between $(m_{\overline{D}}^2)_{23}$
and $(m_{L}^2)_{23}$. The non-trivial structure in the right-handed
neutrino mass matrix may dilute the correlation as in
Eq.~(\ref{sfermionmass}).  However, if the right-handed neutrino masses are
hierarchical, the correlation is expected to be good, as will be shown.

We use two-generation model for simplicity \cite{Ellis:2001xt},
ignoring the first generation. Here, we adopt the parameterization for
the neutrino sector by Casas and Ibarra \cite{ci}.  In this
parametrization the neutrino sector can be parametrized by the
left-handed ($m_{\nu_i}$) and right-handed neutrino masses ($M_{i}$),
the MNS matrix with the Majorana phases ($X$), and a complex orthogonal
matrix ($R$) as
\begin{eqnarray}
U_{ik}^\star f_{\nu_k}{\rm e}^{-i\varphi_{\nu_k}} W_{kj}^\star 
{\rm e}^{-i\overline{\varphi}_{\nu_j}} 
&=&
\frac{1}{\langle H_f\rangle}
X_{ik}^\star \sqrt{m_{\nu_k}}R_{kj}\sqrt{M_{j}}.
\label{casas}
\end{eqnarray}
Using this formula, $(m_{\overline{D}}^2)_{23}$ and $(m_{{L}}^2)_{23}$ 
are  given as 
\begin{eqnarray}
(m_{\overline{D}}^2)_{23} 
&=&
\frac{1}{2(4\pi)^2}
{\rm e}^{-i(\varphi_{d_2}-\varphi_{d_3})}
\frac{(3m_0^2+A_0^2)}{\langle H_f\rangle^2} \log\frac{M_G}{M_{GUT}}
\nonumber\\
&&\left((m_{\nu_\mu}+m_{\nu_\tau})(M_2-M_3) \cos 2\theta_r
+(m_{\nu_\mu}-m_{\nu_\tau})(M_2+M_3) \cosh 2\theta_i\right.
\nonumber\\
&&\left.
-2i\sqrt{m_{\nu_\mu}m_{\nu_\tau}}
((M_2-M_3) \sin\phi \sin 2\theta_r
-(M_2+M_3) \cos\phi \sinh 2\theta_i\right),
\label{md}
\\
(m_{{L}}^2)_{23} 
&=&
\frac{1}{2(4\pi)^2}
\frac{(3m_0^2+A_0^2)}{\langle H_f\rangle^2} 
\nonumber\\
&&\left((m_{\nu_\mu}+m_{\nu_\tau})(\overline{M}_2-\overline{M}_3) \cos 2\theta_r
+(m_{\nu_\mu}-m_{\nu_\tau})(\overline{M}_2+\overline{M}_3) \cosh 2\theta_i\right.
\nonumber\\
&&\left.
-2i\sqrt{m_{\nu_\mu}m_{\nu_\tau}}
((\overline{M}_2-\overline{M}_3) \sin\phi \sin 2\theta_r
-(\overline{M}_2+\overline{M}_3) \cos\phi \sinh 2\theta_i\right)
\label{ml}
\end{eqnarray}
with $\overline{M}_i=M_i \log M_G/M_i$. Here, we use
\begin{eqnarray}
R
&=&
\left(
\begin{array}{cc}
\cos(\theta_r+i \theta_i)
&\sin(\theta_r+i \theta_i)\\
-\sin(\theta_r+i \theta_i)
&\cos(\theta_r+i \theta_i)
\end{array}
\right),
\\
X&=&
\left(
\begin{array}{cc}
1/\sqrt{2}&1/\sqrt{2}\\
-1/\sqrt{2}&1/\sqrt{2}
\end{array}
\right)
\left(
\begin{array}{cc}
{\rm e}^{\phi_M}&\\
&1
\end{array}
\right),
\end{eqnarray}
assuming a maximal mixing for the atmospheric neutrino. $\phi_M$
is the Majorana phase for the light neutrinos. If the right-handed neutrino
masses are hierarchical $(M_3\gg M_2)$, the correlation is good as 
\begin{eqnarray}
\frac{(m_{\overline{D}}^2)_{23}}{(m_{{L}}^2)_{23}} 
&\simeq&
{\rm e}^{-i(\varphi_{d_2}-\varphi_{d_3})}
\frac{\log\frac{M_G}{M_{GUT}}}{\log\frac{M_G}{M_3}}.
\end{eqnarray}

Now we showed that ${(m_{\overline{D}}^2)_{23}}$ and
${(m_{{L}}^2)_{23}}$ are correlated when the right-handed neutrino
masses are hierarchical. We will present the effects of the
right-handed neutrinos in the flavor physics, and the correlations
among the observables. In the following evaluation, we assume 
\begin{eqnarray}
1)~m_{\nu_\tau} &\gg& m_{\nu_\mu}, m_{\nu_e}, \nonumber\\
2)~M_3&\gg& M_2,~M_1,\nonumber\\
3)~W &=& {\bf 1}.
\label{murayama}
\end{eqnarray}
The reason is following. First, the assumption 1) is for simplicity.
Second, ${(m_{\overline{D}}^2)_{23}}$ and ${(m_{{L}}^2)_{23}}$ is 
correlated on the assumption 2), as mentioned above. Furthermore, when
the $M_2$ is comparable to $M_3$, the non-negligible muon-neutrino
Yukawa coupling may lead to large $K^0$-$\bar{K}^0$ mixing
\cite{moroi}\cite{susygutb}\cite{gotoshimizu}\cite{gotoshimizu2} 
and $\mu\rightarrow e \gamma$ \cite{hn} through
$(m_{\overline{D}}^2)_{12}$ and $(m_{L}^2)_{12}$, which are enlarged
by the large solar-neutrino angle.  In some cases, these constraints
are so severe to the SUSY GUT with right-handed neutrinos. Third, $U$
is the MNS matrix on the assumption 3), and this is also for
simplicity. In principle, the non-trivial structure in the
right-handed neutrino mass matrix, which means $R\ne {\bf 1}$ in
Eq.~(\ref{casas}), may enhance the off-diagonal components
${(m_{\overline{D}}^2)_{23}}$ and ${(m_{{L}}^2)_{23}}$ through the
$\cosh 2\theta_i$ and $\sinh 2\theta_i$ in
Eqs.~(\ref{md},\ref{ml}).\footnote { In the case, the light-neutrino
mass matrix has a fine tuning to derive the observed light-neutrino
masses. } Even in the case $R\ne {\bf 1}$, the correlation between
${(m_{\overline{D}}^2)_{23}}$ and ${(m_{{L}}^2)_{23}}$ is kept good
due to the assumption 2). Then, we ignore the case.

On the assumption (\ref{murayama}), the flavor-violating SUSY-breaking
mass term $(m_{\overline{D}}^2)_{23}$ relevant to the CP asymmetries
in $B_d^0\rightarrow
\phi K_s$ ($S_{\phi K_s}$) and $B_s^0\rightarrow J/\psi \phi$
($S_{J/\psi \phi}$) is
\begin{eqnarray}
(m_{\overline{D}}^2)_{23}  &\simeq&-\frac{2}{(4\pi)^2} 
{\rm e}^{-i(\varphi_{d_2}-\varphi_{d_3})} U_{32} U^\star_{33}
\frac{m_{\nu_\tau} M_N}{\langle H_f \rangle^2}
 (3m_0^2+A_0^2) \log\frac{M_G}{M_{GUT}},
\label{msd}
\end{eqnarray}
($M_N\equiv M_3$), and the term $(m_{{L}}^2)_{23}$ for
$Br(\tau \rightarrow \mu \gamma)$ is \footnote{ When the $M_N\lsim
10^{13}$\,GeV, the tau-neutrino Yukawa coupling is too small and
$Br(\tau \rightarrow \mu \gamma)$ is induced by
$(m_{\overline{E}}^2)_{23}$ since it is suppressed by small CKM
elements $V_{32}$ \cite{sglfv}.}
\begin{eqnarray}
(m_{{L}}^2)_{23}  &\simeq&-\frac{2}{(4\pi)^2} 
U_{23} U^\star_{33}
\frac{m_{\nu_\tau} M_N}{\langle H_f \rangle^2}
 (3m_0^2+A_0^2) \log\frac{M_G}{M_{N}}.
\label{msl}
\end{eqnarray}
Here, we use the seesaw formula for the tau-neutrino mass
$m_{\nu_\tau}$. These observables are correlated with each
other up to the CP phase ${\rm
e}^{-i(\varphi_{d_2}-\varphi_{d_3})}$. Furthermore, the muon EDM
$d_\mu$ is proportional to ${\rm Im}~
[(m_{{L}}^2)_{23}(m_{\overline{E}}^2)_{32}]$
\cite{Dimopoulos:1994gj}
where
\begin{eqnarray}
(m_{\overline{E}}^2)_{23}  &\simeq&-\frac{6}{(4\pi)^2} 
{\rm e}^{-i(\varphi_{d_2}-\varphi_{d_3})} 
f_{t}^2V^\star_{32}V_{33}(3m_0^2+ A_0^2)
\log\frac{M_G}{M_{GUT}}.
\end{eqnarray}
Since the imaginary part of $U_{32}U_{33}^\star$ is suppressed due to
small $U_{13}$, the muon EDM is also correlated with the deviation
$S_{\phi K_s}$ and $S_{J/\psi \phi}$, more directly.\footnote{Even in
the MSSM with the right-handed neutrinos, the non-degeneracy of the
right-handed neutrino masses generates the muon EDM
\cite{edmnu}. However, it is proportional to $M_2 M_3\log(M_3/M_2)$,
and the hierarchical right-handed neutrino masses suppress the
effect.}

First, let us explain qualitative behavior of $S_{\phi K_s}$ in the
SUSY GUT with the right-handed neutrinos, and show the correlation
with $Br(\tau\rightarrow\mu\gamma)$ and $d_\mu$.  This is given as
\begin{eqnarray}
S_{\phi K_s}&=&\frac{2 {\rm Im}[\lambda_{\phi K_s}]}{1+|\lambda_{\phi K_s}|^2},
\end{eqnarray}
where
\begin{eqnarray}
\lambda_{\phi K_s} &=&\frac{q_d}{p_d}
\frac{{\cal{A}}(\overline{B}^0_{d}\rightarrow \phi K_s)}{{\cal{A}}({B}^0_{d}\rightarrow \phi K_s)}.
\end{eqnarray}
Here, the ratio $q_{d}/p_{d}$ are for the $B_d^0$-${\overline{B}}_d^0$ mixing. The
explicit formula for the SM and SUSY contributions are given in
Ref.~\cite{hlm}.

If $(m_{\tilde{D}}^2)_{13}$ is not negligible, the SUSY contribution
may change the $B_d^0$-${\overline{B}}_d^0$ mixing from the SM
prediction.  However, the mixing measured in $B_d^0\rightarrow J/\psi
K_S$ is consistent with the SM prediction. Thus, $q_{d}/p_{d}$ should
be dominated by the SM contribution, and it is almost given by
$e^{2i\beta}$.  On the other hand, $\overline{B}^0_{d}\rightarrow
\phi K_s$ is a radiative process, and then the amplitude is
sensitive to the QCD penguin and box contributions with
$(m_{\overline{D}}^2)_{23}$ inserted \cite{Moroi:2000tk}.  In particular,
the QCD penguin contribution with $(m_{\overline{D}}^2)_{23}$ inserted is
proportional to 
$\tan\beta_H(=\langle H_f \rangle/\langle \overline{H}_f \rangle)$,
 and may be dominant for large
$\tan\beta_H$. Since $Br(\tau\rightarrow\mu\gamma)$ and $d_\mu$ are
also proportional to $\tan^2\beta_H$ and $\tan\beta_H$, respectively,
for large $\tan\beta_H$, they are strongly correlated with $S_{\phi
K_S}$.

\begin{figure}
\centerline{
\epsfxsize = 0.5\textwidth \epsffile{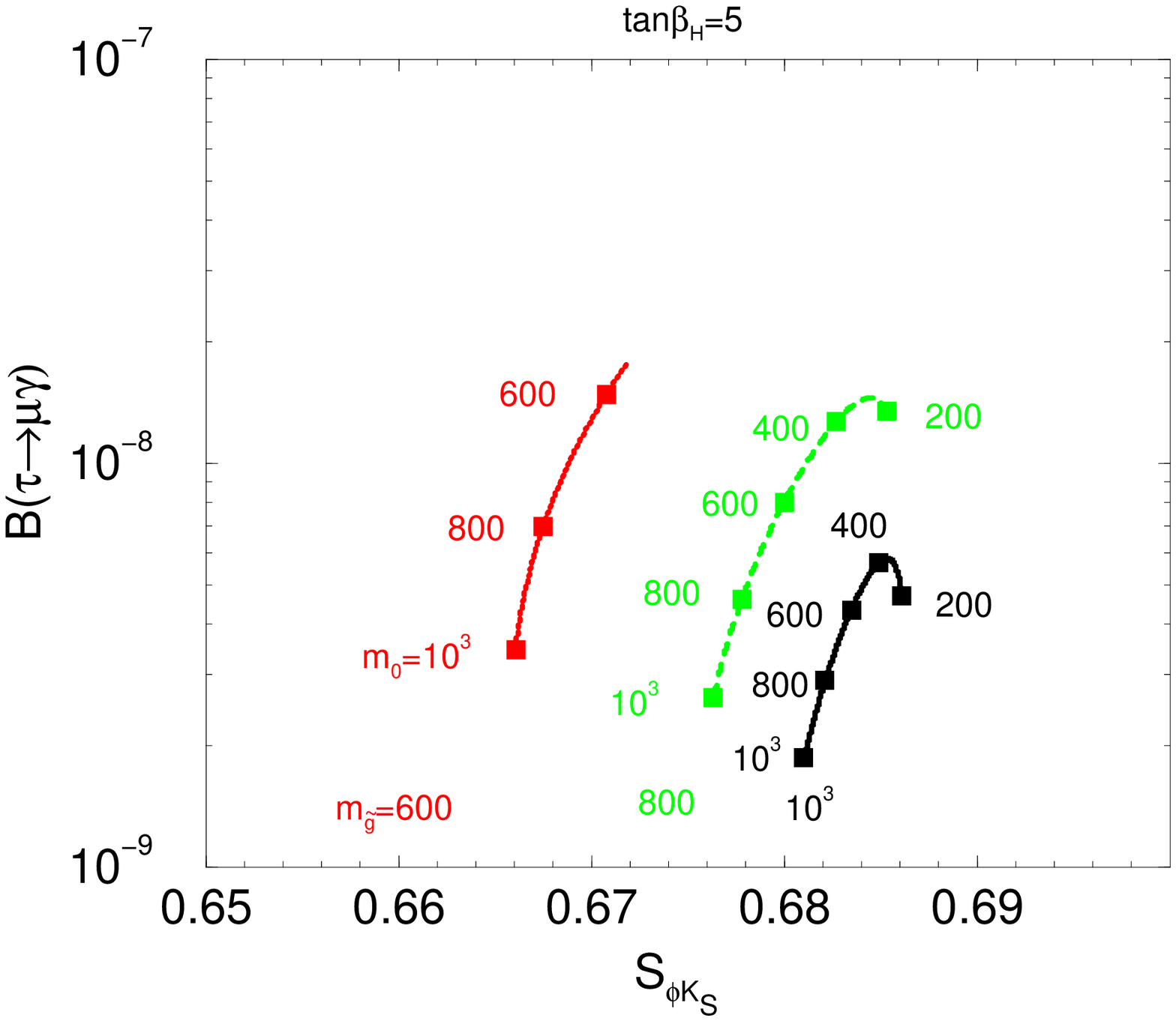} 
\hfill 
\epsfxsize = 0.5\textwidth \epsffile{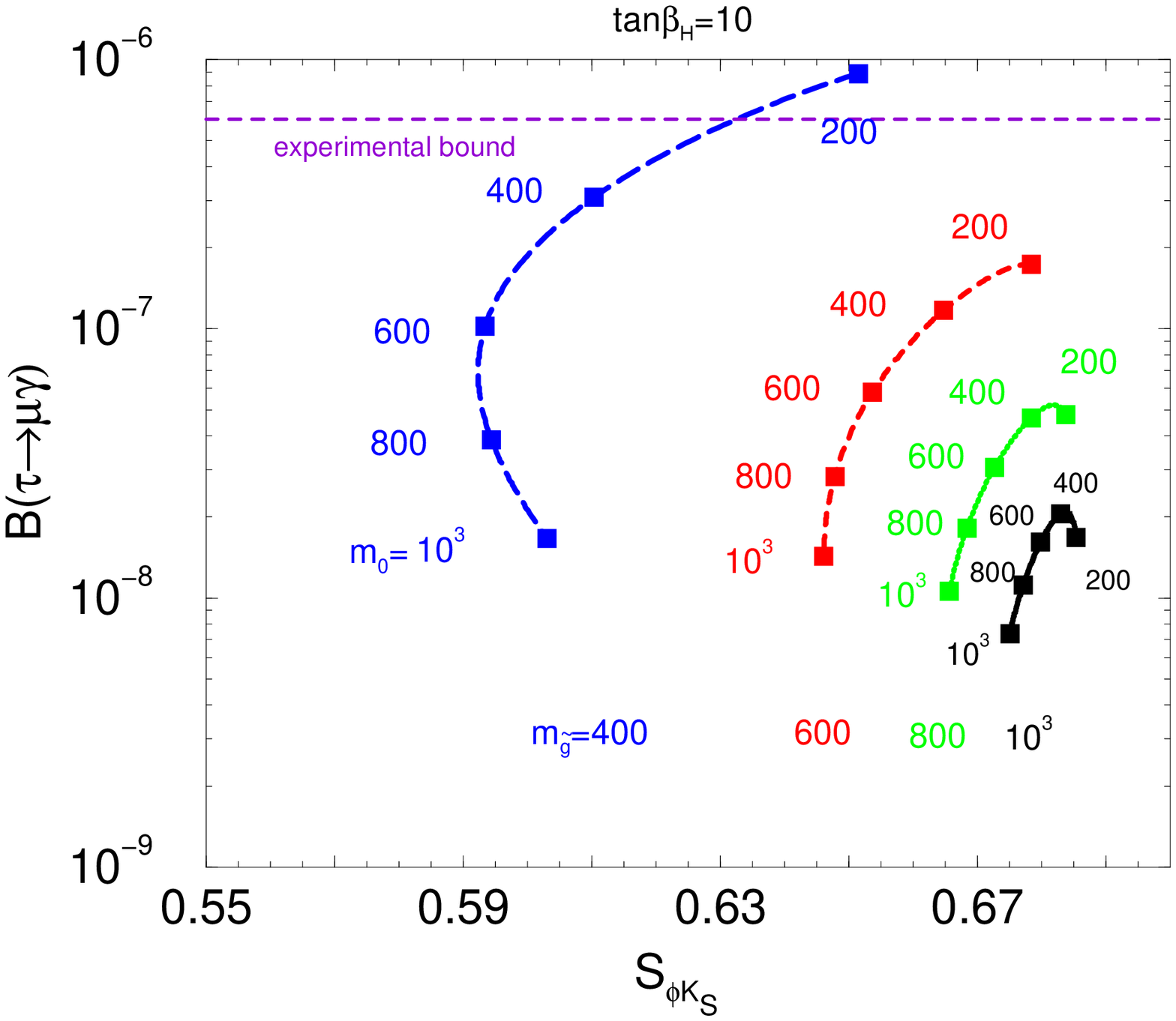} 
}
\centerline{
\epsfxsize = 0.5\textwidth \epsffile{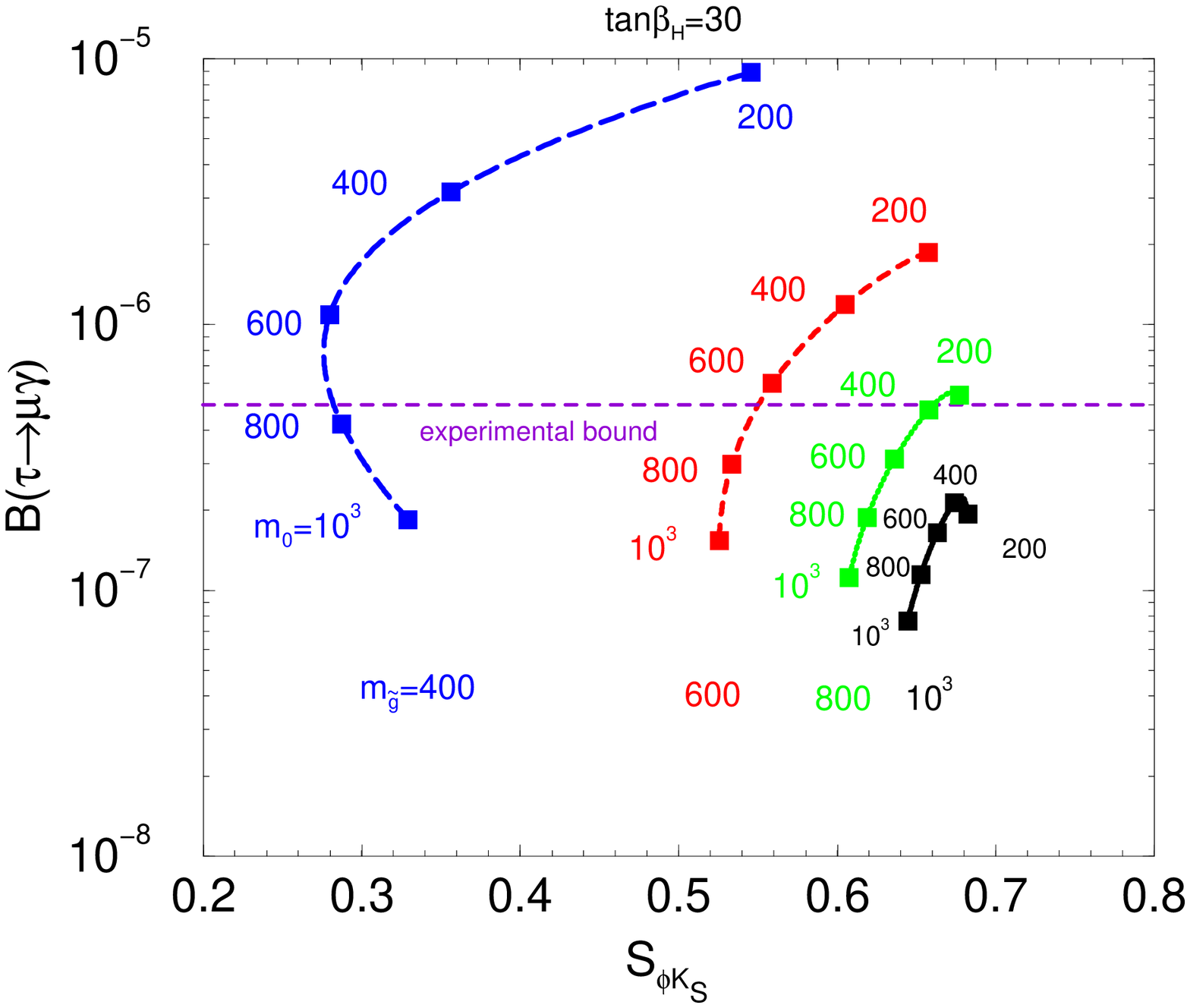} 
}
\vspace*{-5mm}
\caption{
$Br(\tau\rightarrow\mu\gamma)$ as a function of $S_{\phi K_S}$ for
fixed gluino masses $m_{\tilde{g}}=400$, 600, 800,
and 1000\,GeV. $\tan\beta_H$ is 5, 10, and 30. Also,
200\,GeV$<m_0<$1\,TeV, $A_0=0$, $m_{\nu_\tau}=5\times10^{-2}$\,eV,
$M_{N}=5\times 10^{14}$\,GeV,
and $U_{32}=1/\sqrt{2}$. $(\varphi_{d_2}-\varphi_{d_3})$ is taken for the
deviation of $S_{\phi K_S}$ from the SM prediction to be maximum. The
constraints from $b\rightarrow s\gamma$ and the light-Higgs  mass
are imposed.}
\label{fig1}
\end{figure}

In Fig.~(\ref{fig1}) we show $Br(\tau\rightarrow\mu\gamma)$ as a
function of $S_{\phi K_S}$ for fixed gluino masses
$m_{\tilde{g}}$. $\tan\beta_H$ is 5, 10, and 30. Also,
200\,GeV$<m_0<$1\,TeV, $A_0=0$, $m_{\nu_\tau}=5\times10^{-2}$\,eV,
$M_N=5\times 10^{14}$\,GeV, and $U_{32}=1/\sqrt{2}$. The phase
${\rm e}^{i(\varphi_{d_2}-\varphi_{d_3})}$ is taken for the deviation
of $S_{\phi K_S}$ from the SM prediction to be maximum, and the
constraints from $b\rightarrow s\gamma$ and the light-Higgs  mass
are imposed. Here, the matrix element of chromomagnitic moment is
\begin{eqnarray}
\langle \phi K_S|\frac{g_s}{8\pi^2}m_b(\bar{s}_i \sigma^{\mu\nu}T^a_{ij}P_Rb_j)
G^a_{\mu\nu}| \overline{B}_d\rangle &=&
\kappa_{DM} \frac{4\alpha_s}{9\pi}(\epsilon_\phi p_B)f_\phi m_\phi^2 F_+(m_\phi^2),
\end{eqnarray}
and we take $\kappa_{DM}=-1.1$ \cite{hlm}, which is the value in the
heavy-quark effective theory. The deviation is maximized when
$m_{\tilde{g}}$ is lighter and $m_0$ is comparable to $m_{\tilde{g}}$.
While larger $\tan\beta_H$ enhances the deviation of $S_{\phi K_s}$
from the SM prediction ($\sim 0.7$), it is bounded by the constraint from
$Br(\tau\rightarrow\mu \gamma)$.

\begin{figure}
\centerline{
\epsfxsize = 0.5\textwidth \epsffile{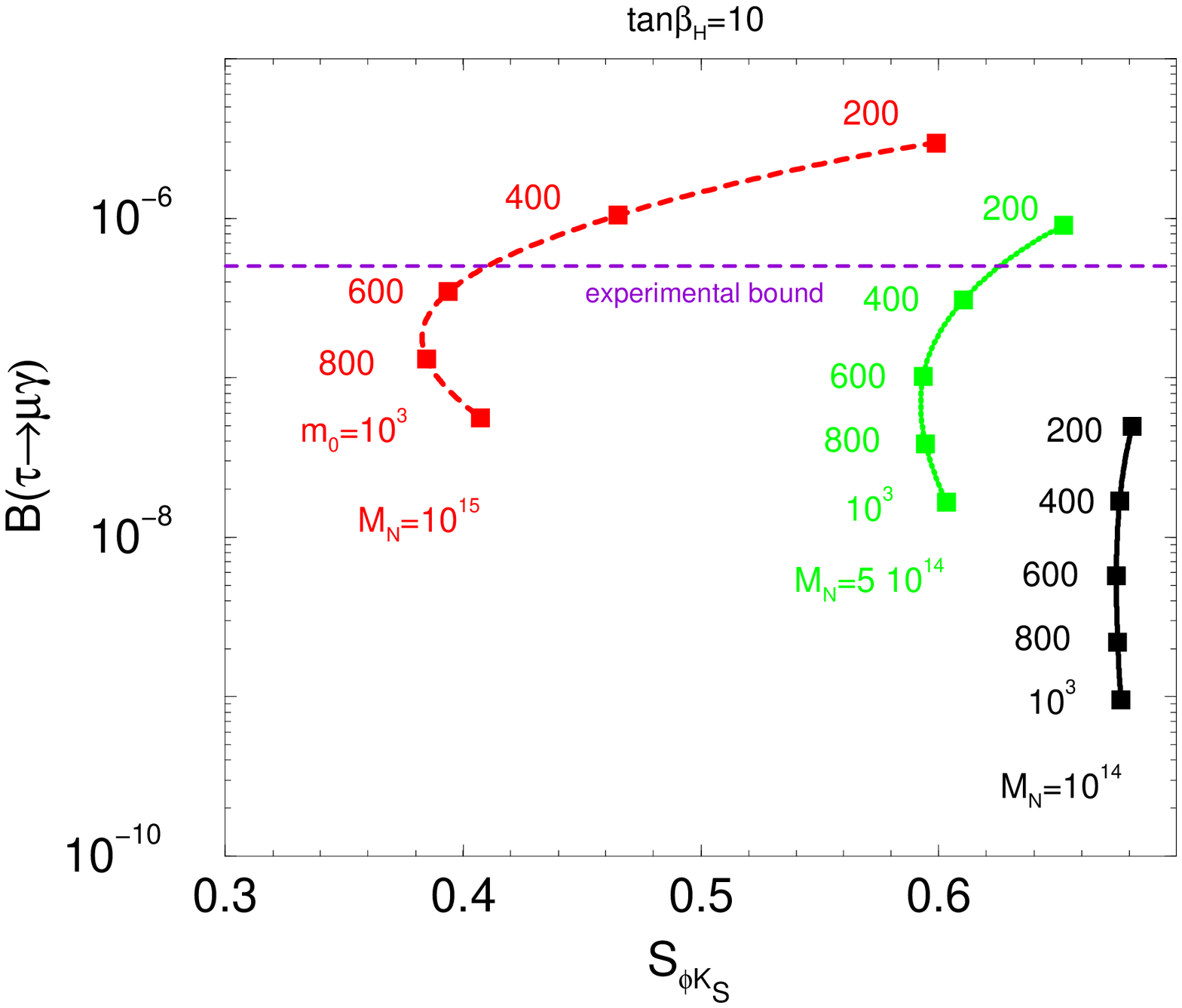} 
\hfill 
\epsfxsize = 0.5\textwidth \epsffile{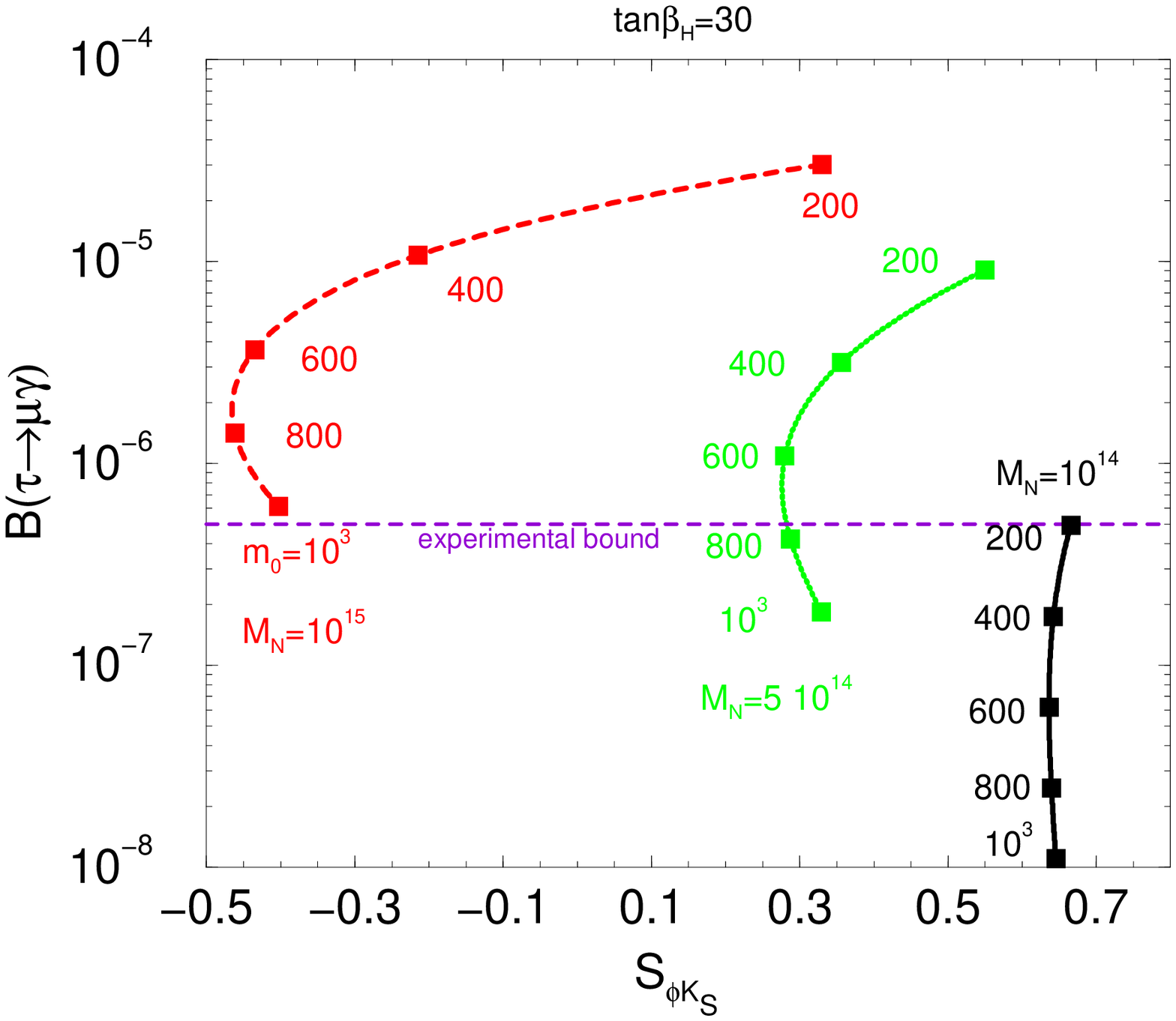} 
}
\vspace*{-5mm}
\caption{
$Br(\tau\rightarrow\mu\gamma)$ as a function of $S_{\phi K_S}$ for
$M_{N}=10^{14}$, $5\times 10^{14}$, and $10^{15}$\,GeV. $\tan\beta_H$ is
10 and 30. Also, 200\,GeV$<m_0<$1\,TeV,
$m_{\tilde{g}}=400$\,GeV, $A_0=0$, $m_{\nu_\tau}=5\times10^{-2}$\,eV,
and $U_{32}=1/\sqrt{2}$. $(\varphi_{d_2}-\varphi_{d_3})$ is taken for the
deviation to be maximum. The constraints from $b\rightarrow s\gamma$
and the light-Higgs mass are imposed.}
\label{fig2}
\end{figure}

Fig.~(\ref{fig2}) is $Br(\tau\rightarrow\mu\gamma)$ as a function of
$S_{\phi K_S}$ for fixed $M_{N}$. Here, $m_{\tilde{g}}=400$\,GeV,
and the other parameters are the same as in Fig.~(\ref{fig1}). From
Eqs.~(\ref{msd}, \ref{msl}), $Br(\tau\rightarrow\mu\gamma)$ and the
SUSY contribution to $\overline{B}_d\rightarrow \phi K_s$ are roughly
proportional to $M_{N}^2$ and $M_{N}$, respectively. When
$M_{N}$ is larger than $10^{15}$\,GeV, the tau-neutrino Yukawa
coupling is larger than 2.7 for $\tan\beta_H=10,30$ and 3.4 for
$\tan\beta_H=5$ at the gravitational scale. From this figure, the
$S_{\phi K_S}$ is limited the bound on $Br(\tau\rightarrow
\mu\gamma)$, and it should be larger than $0.4$ ($-0.3$) for 
$\tan\beta_H=10(30)$ when $M_N$ is smaller than $10^{15}$\,GeV 
and $\kappa_{DM}=-1.1$.

The current bound on $Br(\tau\rightarrow\mu\gamma)$ is $5\times
10^{-7}$ by the Belle experiment \cite{taumug}. The Belle experiment
is expected to obtain about 300\,${\mathrm fb}^{-1}$ data two year after, and 
the bound for $Br(\tau\rightarrow \mu
\gamma)$ is expected to be improved to be $6\times 10^{-8}$. 
Also, if the super $B$ factory starts, it can reach to $2\times
10^{-8}$. Therefore, search for $\tau\rightarrow\mu\gamma$ combined
with measurement of $S_{\phi K_s}$ is expected to play an important
role for consistency check of the SUSY SU(5) GUT with the right-handed
neutrinos.

\begin{figure}
\centerline{
\epsfxsize = 0.5\textwidth \epsffile{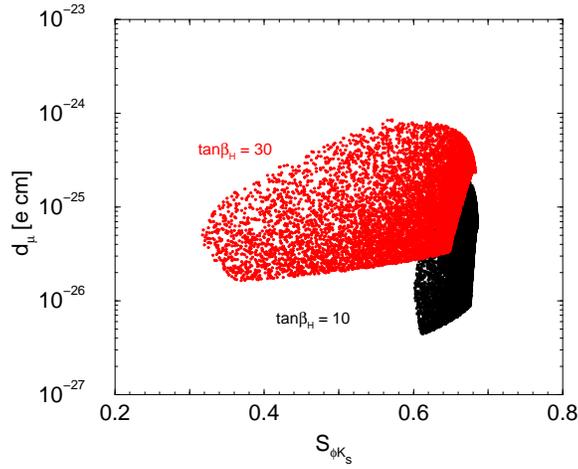} 
}
\vspace*{-5mm}
\caption{
$d_\mu$ and $S_{\phi K_s}$ for $\tan\beta_H=10$ and 30. Here, we take
200\,GeV$<m_0<$1\,TeV and 400\,GeV$<m_{\tilde{g}}<$1\,TeV, $A_0=0$,
$m_{\nu_\tau}=5\times10^{-2}$\,eV, $m_{N_\tau}=5\times 10^{14}$\,GeV,
and $U_{32}=1/\sqrt{2}$. $(\varphi_{d_2}-\varphi_{d_3})$ is taken for the
deviation to be maximum. The constraints from $b\rightarrow s\gamma$,
$\tau\rightarrow \mu\gamma$, and the light-Higgs mass are
imposed.}
\label{fig3}
\end{figure}

As mentioned above, the  muon EDM is also strongly correlated with
$S_{\phi K_s}$.  In Fig.~(\ref{fig3}) we show $d_\mu$ and $S_{\phi
K_s}$ for $\tan\beta_H=10$ and 30 and 200\,GeV$<m_0<$1\,TeV and
400\,GeV$<m_{\tilde{g}}<$1\,TeV.  The other parameters are the same as in
Fig.~(\ref{fig1}). The constraint from $\tau\rightarrow \mu\gamma$ is
imposed in addition to ones from $b\rightarrow s\gamma$ and 
the light-Higgs mass.
From this figure, the sizable deviation of $S_{\phi K_s}$ means
$d_\mu\sim 10^{-(24-25)}e$cm, which might be accessible in the future
muon EDM measurement \cite{PRISM}.

The mixing-induced CP asymmetry $(A_{bsg})$ in $B_d^0\rightarrow
M_s\gamma$ has also a  correlation with $S_{\phi K_s}$ since the
process is induced by the electromagnetic dipole operators and it is
also sensitive to $\tan\beta_H$. The CP asymmetry comes from the
interference with the electromagnetic dipole operators for the
right-handed $c_7$ and left-handed bottom quark $c_7'$ as
\cite{Atwood:1997zr}
\begin{eqnarray}
A_{bsg}
&=&
\frac{2 {\rm Im} [(q_d/p_d) c_7 c_7']}{|c_7|^2+|c_7'|^2}.
\end{eqnarray}
While $c_7'$ is suppressed by $m_s/m_b$ in the SM, non-vanishing
$(m_{\overline{D}}^2)_{23}$ may induces $c_7'$, and $A_{bsg}$ may be
sizable in the SUSY SU(5) GUT with the right-handed neutrino
\cite{gotoshimizu}\cite{gotoshimizu2}. The explicit formula is found
in Ref.~\cite{gotoshimizu2}.  The correlation between $A_{bsg}$ and
$S_{\phi K_S}$ is not necessary for consistency check of the SUSY
SU(5) GUT with the right-handed neutrinos, however, it is important to
prove the origin of the CP and flavor violation, that is
$(m_{\overline{D}}^2)_{23}$.

\begin{figure}
\centerline{
\epsfxsize = 0.5\textwidth \epsffile{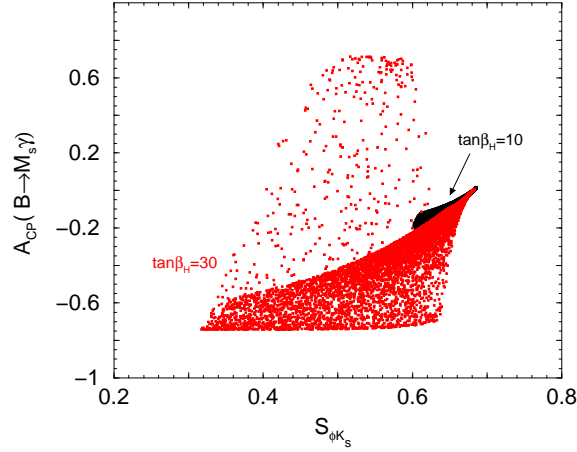} 
}
\vspace*{-5mm}
\caption{
$A_{bsg}$ and $S_{\phi K_s}$ for $\tan\beta_H=10$ and 30. Other
parameters are the same as in Fig.~(\ref{fig3}).}
\label{fig4}
\end{figure}

In Fig~(\ref{fig4}), we show $A_{bsg}$ and $S_{\phi K_S}$ for
$\tan\beta_H=10$ and $30$. The other parameters are the same as in
Fig.~(\ref{fig3}). When $S_{\phi K_S}$ has the sizable deviation from
the SM prediction, $A_{bsg}$ is non-negligible. In the region where $A_{bsg}$
is positive, the sign of $c_7$ is opposite to the SM prediction due the
the SUSY contribution.

Finally, we discuss the CP asymmetry in $B_s^0\rightarrow J/\psi
\phi$.  The asymmetry $S_{J/\psi \phi}$  is given as
\begin{eqnarray}
S_{J/\psi \phi}&=&\frac{2 {\rm
Im}[\lambda_{J/\psi\phi}]}{1+|\lambda_{J/\psi\phi}|^2},
\end{eqnarray}
where
\begin{eqnarray}
\lambda_{J/\psi \phi} &=&\frac{q_s}{p_s}
\frac{{\cal{A}}(\overline{B}^0_{s}\rightarrow J/\psi\phi)}{{\cal{A}}({B}^0_{s}\rightarrow J/\psi\phi)}.
\end{eqnarray}
Here, the ratio $q_{s}/p_{s}$ is for the $B_s^0$-$\overline{B}_s^0$ mixing. Since the
transition $b\rightarrow c\bar{c} s$ is dominated by the tree-level
contribution, the CP asymmetry is almost determined by the
mixing-induced one of $B_s^0$. However, it is tiny in the SM
($\sim -0.03$) since the members of the processes are only two
generations \cite{bs}. Thus, the new physics contribution may dominate
it \cite{jpsiphi}. In particular, $B_s^0\rightarrow J/\psi(\rightarrow
l^+l^-)\phi (\rightarrow K^+K^-)$ is interesting for the $B$ physics
experiments in the hadron colliders \cite{lhc}. If
$(m_{\overline{D}}^2)_{23}$ is non-zero, the box diagram with
$(m_{\overline{D}}^2)_{23}$ inserted may dominate. Since the SUSY
contribution is insensitive to $\tan\beta_H$, the measurement of 
$S_{J/\psi \phi}$ is complimentary to one of $S_{\phi K_s}$.

\begin{figure}
\centerline{
\epsfxsize = 0.5\textwidth \epsffile{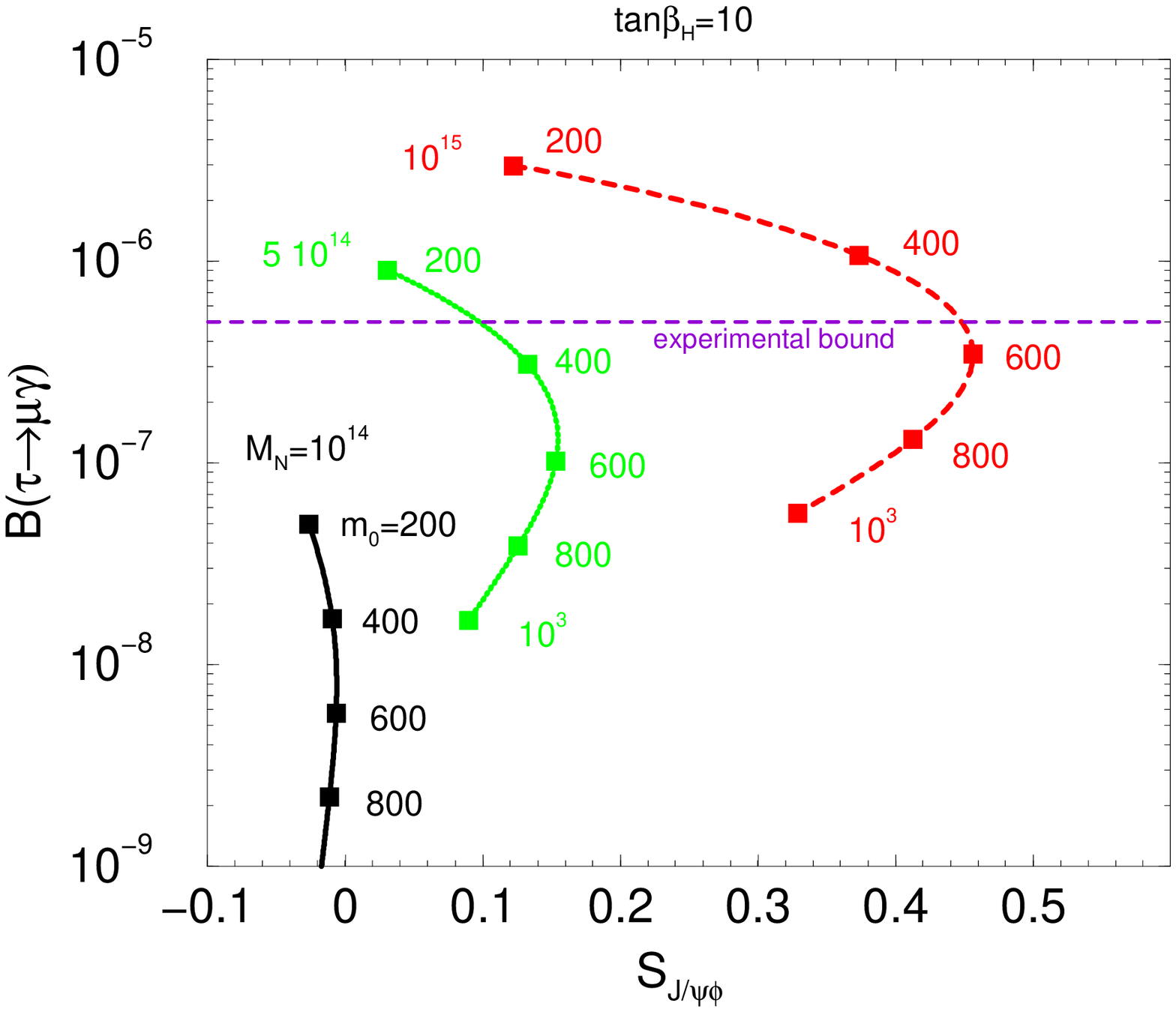} 
\hfill 
\epsfxsize = 0.5\textwidth \epsffile{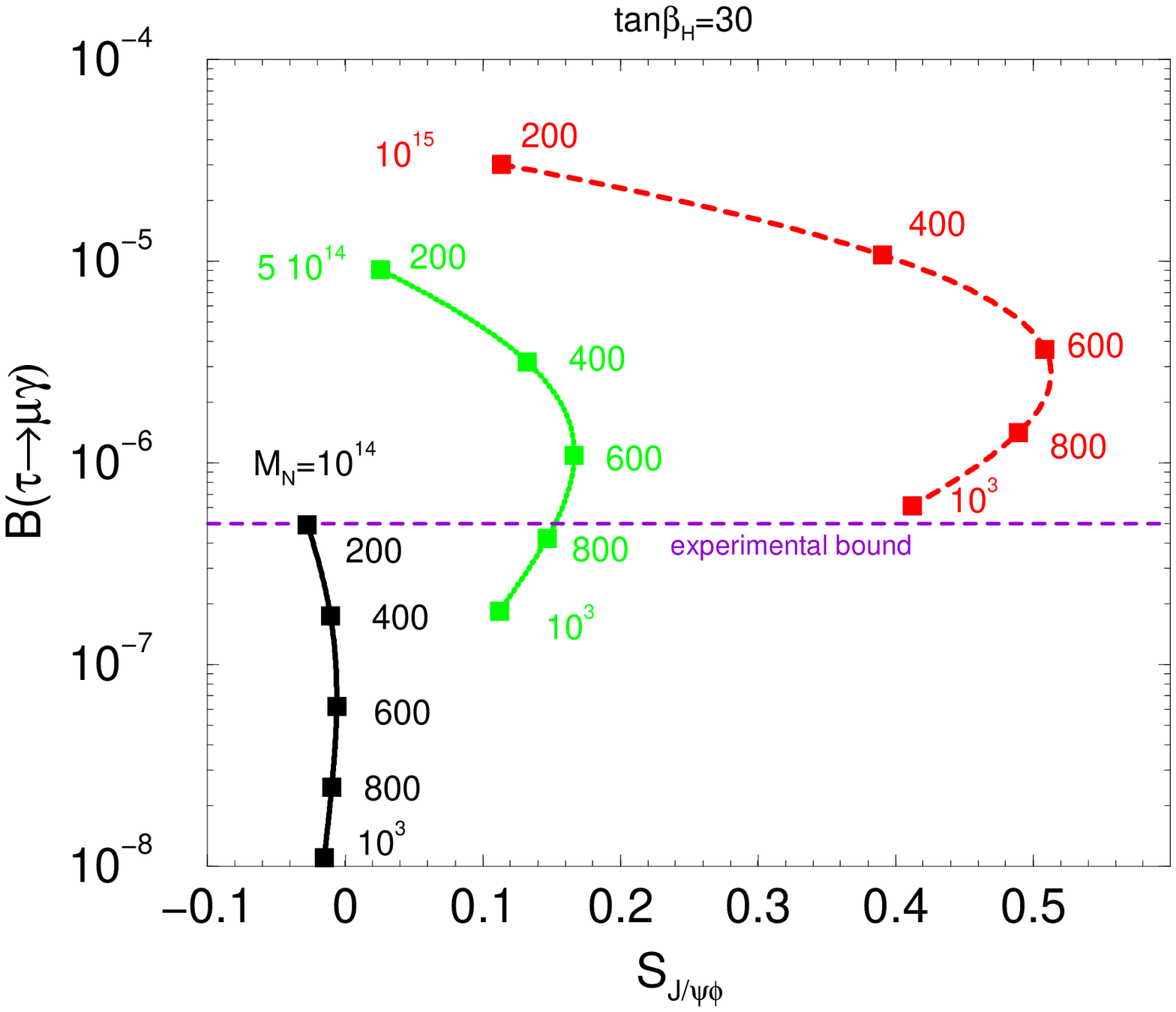} 
}
\vspace*{-5mm}
\caption{
$Br(\tau\rightarrow\mu\gamma)$ as a function of $S_{J/\psi\phi}$
for $M_{N}=10^{14}$, $5\times 10^{14}$, and
$10^{15}$\,GeV. $\tan\beta_H$ is 10  and 30. Other parameters
are the same as in Fig.~(\ref{fig2}).}
\label{fig5}
\end{figure}

In Fig.~(\ref{fig5}) we show $Br(\tau\rightarrow\mu\gamma)$ as a
function of $S_{J/\psi\phi}$ for $M_{N}=10^{14}$, $5\times
10^{14}$, and $10^{15}$\,GeV. Here, $\tan\beta_H$ is 10 and 30. Other
parameters are the same as in Fig.~(\ref{fig2}). Significant
dependence of $S_{J/\psi\phi}$ on $\tan\beta_H$ does not exist as
expected.  $S_{J/\psi \phi}$ can reach to $0.5$ for $M_N<10^{15}$\,GeV.

In Summary, we study the CP asymmetries in $B_d^0\rightarrow \phi K_s$
and $B_s^0\rightarrow J/\psi \phi$, and show the correlations with
$Br(\tau\rightarrow\mu\gamma)$, the muon EDM, and the
mixing-induced CP violation in $B_d^0\rightarrow M^0_s \gamma$ in the SUSY
SU(5) GUT with the right-handed neutrinos.  These observables have a
common origin in this model, which is induced by the tau-neutrino
Yukawa coupling, so that they are correlated. The correlation is better
when the right-handed neutrino masses are hierarchical.  
$S_{\phi K_s}$ has a strong correlation with
$Br(\tau\rightarrow\mu\gamma)$, $d_\mu$, and $A_{bsg}$ for large
$\tan\beta_H$. The measurements for the CP and flavor violation in
both hadronic and leptonic sectors are important for consistency check
of the SUSY SU(5) GUT with the right-handed neutrinos.

\underline{Acknowledgments}

This work is supported in part by the Grant-in-Aid for Science
Research, Ministry of Education, Science and Culture, Japan
(No.~14046225 and No.~1313527 for JH). 
 YS thanks T.~Goto, Y.~Okada, T.~Shindou, and M.~Tanaka
for useful discussions, and the work of YS is supported by
JSPS.

\newpage
%
%
\newcommand{\Journal}[4]{{\sl #1} {\bf #2} {(#3)} {#4}}
\newcommand{\APJ}{Ap. J.}
\newcommand{\CJP}{Can. J. Phys.}
\newcommand{\MPL}{Mod. Phys. Lett.}
\newcommand{\NC}{Nuovo Cimento}
\newcommand{\NP}{Nucl. Phys.}
\newcommand{\PL}{Phys. Lett.}
\newcommand{\PR}{Phys. Rev.}
\newcommand{\PRep}{Phys. Rep.}
\newcommand{\PRL}{Phys. Rev. Lett.}
\newcommand{\PTP}{Prog. Theor. Phys.}
\newcommand{\SJNP}{Sov. J. Nucl. Phys.}
\newcommand{\ZP}{Z. Phys.}

\newpage

\end{document}